\documentclass[12pt]{article}
\usepackage{euler}
\usepackage{amsmath}
\usepackage{amssymb}
\usepackage{amstext}
\usepackage{amscd}
\usepackage{amsfonts}
\DeclareMathAlphabet{\EuFrak}{U}{euf}{m}{n}
\DeclareMathAlphabet{\EuScript}{U}{eus}{m}{n}

\title{{\bf Bosonic String and String Field Theory: a solution 
using the holomorphic representation}
\thanks{\it{This work was partially supported by Consejo
Nacional
de Investigaciones Cient\'{\i}ficas and Comisi\'{o}n de
Investigaciones Cient\'{\i}ficas de la Pcia. de Buenos
Aires;
Argentina.}}}
\author{C.G.Bollini and M.C.Rocca\\
Departamento de F\'{\i}sica, Fac. de Ciencias Exactas,\\
Universidad Nacional de La Plata.\\
C.C. 67 (1900) La Plata. Argentina.}

\date{August 19, 2009}

\begin{document}

\maketitle

\vspace{-5mm}

\begin{abstract}

In this paper we show that the 
holomorphic representation 
is\\appropriate for description
in a consistent way string and string field theories,
when the considered number of component fields of
the string field is finite.
A new Lagrangian for the closed string is obtained
and shown to be equivalent 
to Nambu-Goto's Lagrangian.
We give the notion of anti-string,
evaluate the propagator for the string field,
and calculate the convolution of two of them.


\end{abstract}

\newpage

\renewcommand{\theequation}{\arabic{section}.\arabic{equation}}

\section{Introduction}

In a series of papers \cite{tp1,tp2,tp3,tp4,tp5}
we have shown that Ultradistribution theory of 
Sebastiao e Silva  \cite{tp6,tp7,tp8} permits a significant advance in the treatment 
of quantum field theory. In particular, with the use of the 
convolution of Ultradistributions we have shown that it  is possible
to define a general product of distributions ( a product in a ring
with divisors of zero) that sheds new light on  the question of the divergences
in Quantum Field Theory. Furthermore, Ultradistributions of Exponential Type  
(UET) are  adequate to describe
Gamow States and exponentially increasing fields in Quantun 
Field Theory \cite{tp9,tp10,tp11}.

In four recent papers (\cite{tq1,ts2,ts3,ts4})  we have demonstrated
that Ultradistributions of Exponential type
provide an adequate framework 
for a consistent treatment of 
string and string field theories. In particular, a general
state of the closed string is 
represented by UET of compact support,
and as a consequence the string field is a linear combination
of UET of compact support (CUET). Thus a sting field theory result be
a superposition of infinitely many of fields.
The corresponding development is convergent 
due to that the superposition of infinitely many of complex
Dirac's deltas and its derivatives is convergent.

However for experimental purposes is suitable consider the string 
field theory as a superposition of a 
finite but sufficient great number $n$ 
of fields. 

The resultant theory can be described in a simplified way with the
use of the holomorphic representation \cite{tp16}.
In this case we can not consider a superposition 
of infinitely many of fields due to that we can not 
assure the convergence of the corresponding development 
in powers of the variable $z$. 

In this paper we show that holomorphic representation 
provides an adequate method 
for a consistent simplified treatment of 
closed bosonic string. In particular, a general
state of the closed bosonic string is 
represented by an polynomial  function of a given 
number of complex variables.

This paper is organized as follows:
In section 2 we give a new Lagrangian for bosonic string 
and  solve the corresponding Euler-Lagrange's equations
for closed bosonic string.
In section 3 we give a new representation for the states 
of the string using the holomorphic representation.
In section 4 we give  expressions for the field of the string,
the string field propagator and the creation and annihilation
operators of a string and a anti-string.
In section 5, we give expressions for the non-local action of a free string
and a non-local interaction lagrangian for the string field similar 
to  $\lambda{\phi}^4$ in Quantum Field Theory.
Also we show how to evaluate the convolution
of two string field propagators.
In  section 6 we realize  a discussion of the principal results.

\section{The Constraints for a Bosonic String}

\setcounter{equation}{0}

As is known the Nambu-Goto Lagrangian for the bosonic string 
is given by (\cite{tp13})
\begin{equation}
\label{ep4.1}
{\cal L}_{NG}=T\sqrt{({\dot{X}}\cdot X^{'})^{2}-{\dot{X}}^{2}X^{'2}}
\end{equation}
where
\begin{equation}
\label{ep4.2}
\begin{cases}
X_{\mu}=X_{\mu}(\tau,\sigma)\;;\;
{\dot{X}}_{\mu}={\partial}_{\tau}X_{\mu}\;;\;X^{'}_{\mu}={\partial}_{\sigma}X_{\mu}\\
X_{\mu}(\tau,0)=X_{\mu}(\tau,\pi)\\
-\infty<\tau<\infty\;\;;\;\;0\leq\sigma\leq\pi
\end{cases}
\end{equation}
If we use the constraint
\begin{equation}
\label{ep4.3}
(\dot{X}-X^{'})^{2}=0
\end{equation}
we obtain:
\begin{equation}
\label{ep4.4}
{\dot{X}}^{4}+X^{'4}=4(\dot{X}\cdot X^{'})^{2}-2{\dot{X}}^{2}X^{'2}\geq 0
\end{equation}
On the other hand 
\begin{equation}
\label{ep4.5}
({\dot {X}}^{2}-X^{'2})^{2}={\dot{X}}^{4}+X^{'4}-2{\dot{X}}^{2}X^{'2}
\end{equation}
and from (\ref{ep4.4}) we have
\begin{equation}
\label{ep4.6}
4{\cal L}_{BS}^2=T^2({\dot {X}}^{2}-X^{'2})^{2}=
4T^2[(\dot{X}\cdot X^{'})^{2}-{\dot{X}}^{2}X^{'2}]=4
{\cal L}_{NG}^{2}\geq 0
\end{equation}
As a consequence of (\ref{ep4.6}): 
\begin{equation}
\label{ep4.7}
{\cal L}_{NG}=T\sqrt{(\dot{X}\cdot X^{'})^{2}-{\dot{X}}^{2}X^{'2}}=
\frac {T} {2} |{\dot{X}}^{2}-X^{'2}|={\cal L}_{BS}
\end{equation}
We then see that is sufficient to use only one constraint  to obtain the
Lagrangian for a bosonic string theory from the Namb\'u-Goto
Lagrangian. Another constraint from which  (\ref{ep4.6}) follows is
\begin{equation}
\label{ep4.8}
(\dot{X}+X^{'})^{2}=0
\end{equation}
Thus, the problem for the bosonic string reduces to:
\begin{equation}
\label{ep4.9}
\begin{cases}
{\cal L}=\frac {T} {2}|{\dot{X}}^2-X^{'2}|\\
(\dot{X}+X^{'})^2=0\\
X_{\mu}(\tau,0)=X_{\mu}(\tau,\pi)
\end{cases}
\end{equation}
or 
\begin{equation}
\label{ep4.10}
\begin{cases}
{\cal L}=\frac {T} {2}|{\dot{X}}^2-X^{'2}|\\
(\dot{X}-X^{'})^2=0\\
X_{\mu}(\tau,0)=X_{\mu}(\tau,\pi)
\end{cases}
\end{equation}
The Euler-Lagrange equations for (\ref{ep4.9}) and (\ref{ep4.10}) are
respectively:
\[4\delta({\dot{X}}^2-X^{'2})[(\dot{X}\cdot\ddot{X}-X^{'}\cdot{\dot{X}}^{'}){\dot{X}}_{\mu}-
(X^{'}\cdot {\dot{X}}^{'}-X^{'}\cdot X^{''})X^{'}_{\mu}]+\]
\begin{equation}
\label{ep4.11}
Sgn({\dot{X}}^2-X^{'2})({\ddot{X}}-X^{''})+\lambda(\ddot{X}+2{\dot{X}}^{'}+
 X^{''})=0
\end{equation}
\[4\delta({\dot{X}}^2-X^{'2})[(\dot{X}\cdot\ddot{X}-X^{'}\cdot {\dot{X}}^{'}){\dot{X}}_{\mu}-
(X^{'}\cdot {\dot{X}}^{'}-X^{'}\cdot X^{''})X^{'}_{\mu}]+\]
\begin{equation}
\label{ep4.12}
Sgn({\dot{X}}^2-X^{'2})({\ddot{X}}-X^{''})+\lambda(\ddot{X}-2{\dot{X}}^{'}+
 X^{''})=0
\end{equation}
where $\lambda$ is a Lagrange multiplier.\\
Let $X_{\mu}$ be given by:
\begin{equation}
\label{ep4.13}
X_{\mu}=Sgn({\dot{Y}}^2-Y^{'2}) Y_{\mu}
\end{equation}
where
\begin{equation}
\label{ep4.14}
\begin{cases}
Y_{\mu}(\tau,\sigma)=y_{\mu} + l^2 p_{\mu} \tau+\frac {il} {2}
\sum\limits_{n=-\infty\;;\;n\neq 0}^{\infty}
\frac {a_n} {n} e^{-2in(\tau-\sigma)}\\
p^2=0
\end{cases}
\end{equation}
or
\begin{equation}
\label{ep4.15}
\begin{cases}
Y_{\mu}(\tau,\sigma)=y_{\mu} + l^2p_{\mu} \tau+\frac {il} {2}
\sum\limits_{n=-\infty\;;\; n\neq 0}^{\infty}
\frac {{\tilde{a}}_n} {n} e^{-2in(\tau+\sigma)}\\
p^2=0
\end{cases}
\end{equation}
(\ref{ep4.14}) satisfy 
\begin{equation}
\label{ep4.16}
{\dot{Y}}_{\mu}+Y^{'}_{\mu}=p_{\mu}
\end{equation}
and (\ref{ep4.15})
\begin{equation}
\label{ep4.17}
{\dot{Y}}_{\mu}-Y^{'}_{\mu}=p_{\mu}
\end{equation}
For both we have:
\begin{equation}
\label{ep4.18} 
{\dot{X}}^2-X^{'2}={\dot{Y}}^2-Y^{'2}\neq 0
\end{equation}
and then
\begin{equation}
\label{ep4.19}
({\dot{X}}^2-X^{'2})^2=({\dot{Y}}^2-Y^{'2})^2\neq 0
\end{equation}
((\ref{ep4.13}), (\ref{ep4.14})) and ((\ref{ep4.13}), (\ref{ep4.15})) 
are solutions of (\ref{ep4.11}) and (\ref{ep4.12}) respectively. 
To prove this we take into account that for 
((\ref{ep4.13}), (\ref{ep4.14})) we have
\begin{equation}
\label{ep4.20}
{\ddot{X}}_{\mu}=-{\dot{X}}^{'}=X^{''}
\end{equation}
and for ((\ref{ep4.13}), (\ref{ep4.15}))
\begin{equation}
\label{ep4.21}
{\ddot{X}}_{\mu}={\dot{X}}^{'}=X^{''}
\end{equation}
To quantum level we have respectively for (\ref{ep4.14}) and (\ref{ep4.15}):
\begin{equation}
\label{ep4.22}
\begin{cases}
Y_{\mu}(\tau,\sigma)=y_{\mu} + l^2 p_{\mu} \tau+\frac {il} {2}
\sum\limits_{n=-\infty\;;\;n\neq 0}^{\infty}
\frac {a_{n\mu}} {n} e^{-2in(\tau-\sigma)}\\
p^2|\phi>=0
\end{cases}
\end{equation}
and
\begin{equation}
\label{ep4.23}
\begin{cases}
Y_{\mu}(\tau,\sigma)=y_{\mu} + l^2p_{\mu} \tau+\frac {il} {2}
\sum\limits_{n=-\infty\;;\; n\neq 0}^{\infty}
\frac {{\tilde{a}}_{n\mu}} {n} e^{-2in(\tau+\sigma)}\\
p^2|\phi>=0
\end{cases}
\end{equation}
where $|\Phi>$ is the physical state of string.\\
In terms of creation and annihilation operators we have for
(\ref{ep4.22}) and (\ref{ep4.23}):
\begin{equation}
\label{ep4.24}
\begin{cases}
Y_{\mu}(\tau,\sigma)=y_{\mu} + l^2 p_{\mu} \tau+\frac {il} {2}
\sum\limits_{n>0}
\frac {b_{n\mu}} {\sqrt{n}} e^{-2in(\tau-\sigma)}-
\frac {b^+_{n\mu}} {\sqrt{n}} e^{2in(\tau-\sigma)}\\
p^2|\phi>=0
\end{cases}
\end{equation}
\begin{equation}
\label{ep4.25}
\begin{cases}
Y_{\mu}(\tau,\sigma)=y_{\mu} + l^2p_{\mu} \tau+\frac {il} {2}
\sum\limits_{n>0}
\frac {{\tilde{b}}_{n\mu}} {\sqrt{n}} e^{-2in(\tau+\sigma)}-
\frac {{\tilde{b}}^+_{n\mu}} {\sqrt{n}} e^{-2in(\tau+\sigma)}\\
p^2|\phi>=0
\end{cases}
\end{equation}
where:
\begin{equation}
\label{ep4.26}
[b_{\mu m},b^+_{\nu n}]={\eta}_{\mu\nu}
{\delta}_{mn}
\end{equation}
\begin{equation}
\label{ep4.27}
[{\tilde{b}}_{\mu m},{\tilde{b}}^+_{\nu n}]={\eta}_{\mu\nu}
{\delta}_{mn}
\end{equation}
A general state of the string can be written as:
\[|\phi>=[a_0(p)+a^{i_1}_{\mu_1}(p)b^{+\mu_1}_{i_1}+
a^{i_1 i_2}_{\mu_1\mu_2}(p)b^{+\mu_1}_{i_1}b^{+\mu_2}_{i_2}
+...+...\]
\begin{equation}
\label{ep4.28}
+a^{i_1i_2...i_n}_{\mu_1\mu_2...\mu_n}(p)b^{+\mu_1}_{i_1}
b^{+\mu_2}_{\i_2}...b^{+\mu_n}_{i_n}+...+...]
|0>
\end{equation}
or
\[|\phi>=[a_0(p)+a^{i_1}_{\mu_1}(p){\tilde{b}}^{+\mu_1}_{i_1}+
a^{i_1 i_2}_{\mu_1\mu_2}(p){\tilde{b}}^{+\mu_1}_{i_1}{\tilde{b}}^{+\mu_2}_{i_2}
+...+...\]
\begin{equation}
\label{ep4.29}
+a^{i_1i_2...i_n}_{\mu_1\mu_2...\mu_n}(p){\tilde{b}}^{+\mu_1}_{i_1}
{\tilde{b}}^{+\mu_2}_{\i_2}...{\tilde{b}}^{+\mu_n}_{i_n}+...+...]
|0>
\end{equation}
where:
\begin{equation}
\label{ep4.30}
p^2 a^{i_1i_2...i_n}_{\mu_1\mu_2...\mu_n}(p)=0
\end{equation}
It  is immediate to prove that ((\ref{ep4.13}), (\ref{ep4.14})) and 
((\ref{ep4.13}), (\ref{ep4.15})) are solutions  
of Nambu-Goto equations on physical states. 
(Nambu-Goto equations arise from Euler-Lagrange equations 
corresponding to the  Lagrangian
(\ref{ep4.1}), and it is easy to prove that the currently used  solution 
for the  closed string movement 
is not solution of Nambu-Goto equations 
due to the fact that Virasoro operators $L_n$ and ${\tilde{L}}_n$
does not annihilate the physical states for $n<0$ and moreover,
does not form a set of commuting operators).

\section{A representation of  the states of the closed string}

Let ${\large{A}}(\mathbb{R}^2)$ be the complex Euclidean space
defined as (see ref.\cite{tp15} for a \\definition of complex Euclidean space)
\setcounter{equation}{0}
\begin{equation}
\label{ep3.1}
{\large{A}}(\mathbb{R}^2)
=\left\{f(z)/f(z) \;is \;analytic \;\wedge
\frac {i} {2}\int f(z)\overline{f(z)}e^{-z\overline{z}}dz\;d\overline{z}<
\infty\right\}
\end{equation}
supplied with the complex scalar product (see ref.\cite{tp16}):
\begin{equation}
\label{ep3.2}
<f(z),g(z)>=
\frac {i} {2}\int f(z)\overline{g(z)}e^{-z\overline{z}}dz\;d\overline{z}=
\end{equation}
Taking into account that:
\[\frac {i} {2} dz\wedge d\overline{z}=
\frac {i} {2} (dx+idy)\wedge(dx-idy)=\]
\begin{equation}
\label{ep3.3}
\frac {1} {2} dx\wedge dy-dy\wedge dx=
dx\wedge dy
\end{equation}
where $\wedge$ denotes the outer product of two 1-forms,
we have for the scalar product (\ref{ep3.2})
\begin{equation}
\label{ep3.4}
<f(z),g(z)>=\iint\limits_{-\infty}^{\;\;\;\infty}f(z)\overline{g(z)}
e^{-z\overline{z}}dx\;dy=\int\limits_{-\infty}^{\infty}
\int\limits_0^{2\pi}f(z)\overline{g(z)}\rho e^{-{\rho}^2} 
d\theta\;d\rho
\end{equation}
Let ${\large{Z}}(\mathbb{R}^2)$ be the complex
Euclidean space defined as:
\begin{equation}
\label{ep3.5}
{\large{Z}}(\mathbb{R}^2)
=\left\{f(z)/z^p\frac {d^qf(z)} {d^qz}\in{\large{A}}(\mathbb{R}^2)\right\}
\end{equation}
supplied with the scalar product (\ref{ep3.2}), where 
$p$ and $q$ are natural numbers. For $f\in{\large{Z}}(\mathbb{R}^2)$
we have:
\[<\frac {df(z)} {dz},g(z)>=\frac {i} {2} \int \frac {df(z)} {dz} \overline{g(z)}
e^{-z\overline{z}}dz\;d\overline{z}=\]
\begin{equation}
\label{ep3.6}
\frac {i} {2} \int f(z)\overline{zg(z)}e^{-z\overline{z}}
dz\;d\overline{z}=<f(z),zg(z)>
\end{equation}
If we define:
\begin{equation}
\label{ep3.7}
a=\frac {d} {dz}\;\;\;;\;\;\;a^+=z
\end{equation}
we obtain:
\begin{equation}
\label{ep3.8}
[a,a^+]=1
\end{equation}
Representation (\ref{ep3.7}) is called the holomorphic representation
(see ref.\cite{tp16}) for annihilation and creation operators.

The vacuum state annihilated by $d/dz$ is the number $1/\sqrt{\pi}$ and the
orthonormalized states obtained by successive application $z$ to $1/\sqrt{\pi}$ are:
\begin{equation}
\label{ep3.9}
 F_n(z)=\frac {z^n} {\sqrt{\pi\; n!}}
\end{equation}
Using this representation a general state of the string can be written as:
\[\phi(x,\{z\})=a_0(x)+a^{i_1}_{\mu_1}(x)z^{\mu_1}_{i_1}+
a^{i_1 i_2}_{\mu_1\mu_2}(x)z^{\mu_1}_{i_1}z^{\mu_2}_{i_2}
+...+...\]
\begin{equation}
\label{ep3.10}
+a^{i_1i_2...i_n}_{\mu_1\mu_2...\mu_n}(x)z^{\mu_1}_{i_1}
z^{\mu_2}_{\i_2}...z^{\mu_n}_{i_n}
\end{equation}
where $\{z\}$ denotes $(z_{1\mu},z_{2\mu},...,z_{n\mu})$.\\
The functions
$a^{i_1i_2...i_n}_{\mu_1\mu_2...\mu_n}(x)$
are solutions of
\begin{equation}
\label{ep3.11}
\Box a^{i_1i_2...i_n}_{\mu_1\mu_2...\mu_n}(x)=0
\end{equation}

\section{The String Field}

\setcounter{equation}{0}

According to (\ref{ep4.25}), (\ref{ep4.25}) and section 3 the equation for the string field
is given by:
\begin{equation}
\label{ep8.1}
\Box\Phi(x,\{z\})=({\partial}^2_0-{\partial}^2_1-{\partial}^2_2-{\partial}^2_3)
\Phi(x,\{z\})=0
\end{equation}
where $\{z\}$ denotes $(z_{1\mu},z_{2\mu},...,z_{n\mu},...,....)$, and 
$\Phi$ is a analytic function in the set of variables $\{z\}$.
Thus we have:
\[\Phi(x,\{z\})=[A_0(x)+A^{i_1}_{\mu_1}(x)z^{\mu_1}_{i_1}+
A^{i_1 i_2}_{\mu_1\mu_2}(x)z^{\mu_1}_{i_1}z^{\mu_2}_{i_2}
+...+...\]
\begin{equation}
\label{ep8.2}
+A^{i_1i_2...i_n}_{\mu_1\mu_2...\mu_n}(x)z^{\mu_1}_{i_1}
z^{\mu_2}_{\i_2}...z^{\mu_n}_{i_n}
\end{equation}
where the quantum fields 
$A^{i_1i_2...i_n}_{\mu_1\mu_2...\mu_n}(x)$
are solutions of
\begin{equation}
\label{ep8.3}
\Box A^{i_1i_2...i_n}_{\mu_1\mu_2...\mu_n}(x)=0
\end{equation}
The propagator of the string field can be expressed in terms of the propagators 
of the component fields:
\[\Delta(x-x^{'},\{z\},\{{\overline{z}}^{'}\})=\Delta_0(x-x^{'})+\Delta^{i_1j_1}_{\mu_1\mu_2}
(x-x^{'})z_{i_1}^{\mu_1}{\overline{z}}_{j_1}^{'\nu_1}+...+...+\]
\begin{equation}
\label{ep8.4}
\Delta^{i_1...i_nj_1...j_n}_{\mu_1...\mu_n\nu_1...\nu_n}(x-x^{'})
z^{\mu_1}_{i_1}...z^{\mu_n}_{i_n}{\overline{z}}^{'\nu_1}_{j_1}...
{\overline{z}}^{'\nu_n}_{j_n}
\end{equation}
For the fields $A^{i_1i_2...i_n}_{\mu_1\mu_2 ...\mu_n}(x)$ we have:
\begin{equation}
\label{ep8.5}
A^{i_1i_2...i_n}_{\mu_1\mu_2...\mu_n}(x)=\int\limits_{-\infty}^{\infty}
a^{i_1i_2...i_n}_{\mu_1\mu_2...\mu_n}(k)e^{-ik_{\mu}x^{\mu}}+
b^{+i_1i_2...i_n}_{\mu_1\mu_2...\mu_n}(k)e^{ik_{\mu}x^{\mu}}\;
d^3k
\end{equation}

We define the operators of annihilation and creation of a string as:
\[a(k,\{z\})=a_0(k)+a_{\mu_1}^{i_1}(k)z_{i_1}^{\mu_1}+...+...+\]
\begin{equation}
\label{ep8.6}
a_{\mu_1...\mu_n}^{i_1...i_n}(k)z_{i_1}^{\mu_1}...z_{i_n}^{\mu_n}
\end{equation}
\[a^+(k^{'},\{{\overline{z}}^{'}\})=a^+_0(k^{'})+a_{\nu_1}^{+j_1}(k^{'})
{\overline{z}}_{j_1}^{'\nu_1}+...+...+\]
\begin{equation}
\label{ep8.7}
a_{\nu_1...\nu_n}^{+j_1...j_n}(k^{'}){\overline{z}}_{j_1}^{'\nu_1}...{\overline{z}}_{j_n}^{'\nu_n}
\end{equation}
and the annihilation and creation operators for the anti-string 
\[b(k,\{{\overline{z}}\})=b_0(k)+b_{\mu_1}^{i_1}(k){\overline{z}}_{i_1}^{\mu_1}+...+...+\]
\begin{equation}
\label{ep8.8}
b_{\mu_1...\mu_n}^{i_1...i_n}(k){\overline{z}}_{i_1}^{\mu_1}...{\overline{z}}_{i_n}^{\mu_n}
\end{equation}
\[b^+(k^{'},\{z^{'}\})=b^+_0(k^{'})+b_{\nu_1}^{+j_1}(k^{'})z_{j_1}^{'\nu_1}+...+...+\]
\begin{equation}
\label{ep8.9}
b_{\nu_1...\nu_n}^{+j_1...j_n}(k^{'})z_{j_1}^{'\nu_1}...z_{j_n}^{'\nu_n}
\end{equation}
If we define
\begin{equation}
\label{ep8.10}
[a_{\mu_1...\mu_n}^{i_1...i_n}(k),a_{\nu_1..\nu_n}^{+j_1...j_n}(k^{'})]=
f_{\mu_1...\mu_n\nu_1...\nu_n}^{i_1...i_nj_1...j_n}(k)\delta(k-k^{'})
\end{equation}
the commutations relations are
\[[a(k,\{z\}),a^+(k^{'},\{{\overline{z}}^{'}\})]=[f_0(k)+f_{\mu_1\nu_1}^{i_1j_1}(k)
z_{i_1}^{\mu_1}{\overline{z}}_{j_1}^{'\nu_1}+...+...\]
\begin{equation}
\label{ep8.11}
f_{\mu_1...\mu_n\nu_1...\nu_n}^{i_1...i_nj_1...j_n}(k)
z_{i_1}^{\mu_1}...z_{i_n}^{\mu_n}
{\overline{z}}_{j_1}^{'\nu_1}...{\overline{z}}_{j_n}^{'\nu_n}]
\delta(k-k^{'})
\end{equation}
and for the anti-string:
\begin{equation}
\label{ep8.12}
[b_{\mu_1...\mu_n}^{i_1...i_n}(k),b_{\nu_1..\nu_n}^{+j_1...j_n}(k^{'})]=
g_{\mu_1...\mu_n\nu_1...\nu_n}^{i_1...i_nj_1...j_n}(k)\delta(k-k^{'})
\end{equation}
the commutations relations are
\[[b(k,\{\overline{z}\}),b^+(k^{'},\{z^{'}\})]=[g_0(k)+g_{\mu_1\nu_1}^{i_1j_1}(k)
{\overline{z}}_{i_1}^{\mu_1}z_{j_1}^{'\nu_1}+...+...\]
\begin{equation}
\label{ep8.13}
g_{\mu_1...\mu_n\nu_1...\nu_n}^{i_1...i_nj_1...j_n}(k)
{\overline{z}}_{i_1}^{\mu_1}...{\overline{z}}_{i_n}^{\mu_n}
z_{j_1}^{'\nu_1}...z_{j_n}^{'\nu_n}]
\delta(k-k{'})
\end{equation}
With this anihilation and creation operators we can write:
\begin{equation}
\label{ep8.14}
\Phi(x,\{z\})=\int\limits_{-\infty}^{\infty}
a(k,\{z\})e^{-ik_{\mu}x^{\mu}}+
b^+(k\{z\})e^{ik_{\mu}x^{\mu}}\;
d^3k
\end{equation}

\section{The Action for the String Field}

\setcounter{equation}{0}

The action for the free bosonic bradyonic closed string field is:
\begin{equation}
\label{ep9.1}
S_{free}=\frac {i^n} {2^n}
\int\int\limits_{-\infty}^{\infty}
\partial_{\mu}\Phi(x,\{z\})e^{-\{z\}{\cdot}\{\overline{z}\}}
\partial^{\mu}\Phi^+(x,\{z\})\;d^3x\;\{dz\}\;\{d\overline{z}\}
\end{equation}
A possible interaction is given by:
\[S_{int}=\lambda\;\frac {i^n} {2^n}\;\int
\int\limits_{-\infty}^{\infty}
\Phi(x,\{z\})e^{-\{z\}{\cdot}\{\overline{z}\}}
\Phi^+(x,\{z\})e^{-\{\overline{z}\}{\cdot}\{z\}}
\Phi(x,\{z\})\times\]
\begin{equation}
\label{ep9.2}
e^{-\{z\}{\cdot}\{\overline{z}\}}
\Phi^+(x,\{z\})\;
d^3x\;\{dz\}\;\{d\overline{z}\}
\end{equation}
Both, $S_{free}$ and $S_{int}$ are non-local as expected.

The convolution
of two propagators of  the string field is:
\begin{equation}
\label{ep9.6}
\hat{\Delta}(k,\{z_1\},\{{\overline{z}}_2\})\ast
\hat{\Delta}(k,\{z_3\},\{{\overline{z}}_4\})
\end{equation}
where $\ast$ denotes the convolution
of Ultradistributions of Exponential Type  
on the $k$ variable only.
With the use of the result
\begin{equation}
\label{ep9.7}
\frac {1} {\rho}\ast\frac {1} {\rho}=-\pi^2\ln\rho
\end{equation}
($\rho=x_0^2+x_1^2+x_2^2+x_3^2$ in euclidean space)

and
\begin{equation}
\label{ep9.8}
\frac {1} {\rho\pm i0}\ast\frac {1} {\rho\pm i0}=
 \mp i\pi^2\ln(\rho\pm i0)
\end{equation}
($\rho=x_0^2-x_1^2-x_2^2-x_3^2$ in minkowskian space)

the convolution of two string field propagators is finite.

\section{Discussion}

We have shown that holomorphic representation is appropriate for
the description 
in a consistent way string and string field theories.
By means of  a new Lagrangian for the closed string strictly equivalent
to Nambu-Goto Lagrangian we have obtained a movement 
equation for the field of the string and solve it.
We shown that this string field 
is a polynomial in the variables $z$.
We evaluate the propagator for the string field,
and calculate the convolution of two of them, taking
into account that string field theory is a non-local theory.
For practical calculations and experimental results 
we have given expressions that
involve only a finite number of variables.

As a final remark we would like to point out that our formulas
for convolutions follow from  general definitions. They are not
regularized expressions

\end{document}